\documentclass[12pt,titlepage]{utarticle}
\usepackage{amsmath,amssymb,amsthm,pictexwd,graphicx,color}
\usepackage{setspace}
\usepackage[mathscr]{eucal}
\usepackage[colorlinks=true, pdfstartview=FitV, linkcolor=blue, citecolor=blue, urlcolor=blue]{hyperref}
\usepackage{xypic}

\numberwithin{equation}{section}

\def\wdg{{\wedge}}                              % wedge product

\def\BZ{\mathbb{Z}}

\def\BR{\mathbb{R}}
\newtheorem{defn}{Definition}
\newtheorem{thm}{Theorem}

\def\mcA{\mathcal{A}}

\def\mcD{\mathcal{D}}

\def\mcF{\mathcal{F}}

\def\mcT{\mathcal{T}}
\def\mcL{\mathcal{L}}

\def\mcV{\mathcal{V}}

\def\Tr{\mathrm{Tr}}

\def\cl{C\!\ell}
\def\Spin{\mathrm{Spin}}

\def\ie{\textit{i.e.,\ }}

\newcommand{\be}{\begin{equation}}
\newcommand{\ee}{\end{equation}}
\newcommand{\bea}{\begin{eqnarray}}
\newcommand{\eea}{\end{eqnarray}}
\newcommand{\al}{\alpha}

\newcommand{\G}{\Gamma}
\newcommand{\g}{\gamma}

\newcommand{\Om}{\Omega}
\newcommand{\om}{\omega}
\newcommand{\s}{\sigma}

\newcommand{\hlf}{\frac{1}{2}}

\newcommand{\M}{\mathcal{M}}
\newcommand{\non}{\nonumber}

\newcommand{\p}{\partial}

\newcommand{\R}{\mathbb{R}}
\newcommand{\rr}{\rightarrow}
\newcommand{\w}{\wedge}
\newcommand{\Z}{\mathbb{Z}}
\newcommand{\GL}{\operatorname{GL}}

\renewcommand{\O}{\operatorname{O}}

\newcommand{\SL}{\operatorname{SL}}
\newcommand{\SO}{\operatorname{SO}}

\newcommand{\U}{\operatorname{U}}
\newcommand{\lp}{\left(}
\newcommand{\rp}{\right)}
\newcommand{\ls}{\left[}
\newcommand{\rs}{\right]}

\newcommand{\VV}{V\oplus V^\ast}

\begin{document}
\preprint{UTTG--12--07\\ MIFP--07--27\\ }

\title{Ramond-Ramond Fields, Cohomology and Non-Geometric Fluxes}

\author{Aaron Bergman\address{George P. \& Cynthia W. Mitchell Institute for Fundamental Physics\\
             Texas A\&M University\\
             College Station, TX 77843-4242\\
             \email{abergman@physics.tamu.edu}}
             and Daniel Robbins\address{Theory Group, Physics Department\\
             University of Texas at Austin\\
	   Austin, TX 78712\\ {~}\\
	   \email{robbins@zippy.ph.utexas.edu}}}

\Abstract{We consider compactifications of type II string theory in which a
$d$-dimensional torus is fibered over a base $X$.  In string theory, the
transition functions of this fibration need not be simply diffeomorphisms of
$T^d$ but can involve elements of the T-duality group
$\operatorname{Spin}(d,d,\Z)$.  We precisely define the notion of a T-fold
with NSNS flux. Given such a T-fold, we define the $\Z_2$-graded cohomology
theory describing the unquantized RR field strengths and discuss how the
data of a T-fold can be interpreted in terms of generalized NSNS fluxes and
the twisted differential of Shelton-Taylor-Wecht.}

\maketitle
\newpage

\section{Introduction}\label{sec:intro}

Below we present two independent introductions to the material of the paper, one aimed at readers in the physics community, and one aimed at mathematicians.

\subsection{Introduction for physicists}

In an effort to construct vacua of string theory with semi-realistic or perhaps even phenomenologically viable four-dimensional physics, many researchers have considered so-called flux compactifications.  In these constructions one specifies not only the geometry of the internal space $X$ but also some extra data describing the configuration of RR and NSNS field strengths.  For a fixed $X$, these data are typically classified by various cohomology or K-theory groups of $X$.  Physically, including these extra ingredients can be very useful.  They tend to generate potentials for the scalar fields of the theory leading to improved moduli stabilization, spontanteous supersymmetry breaking, the possibility of inflation in some regions of field space, and other phenomena.  Unfortunately, even with fluxes, the structure of the four-dimensional effective potential may still not be as rich as we might like.  For instance we cannot stabilize all moduli perturbatively without appealing to extra effects (from D-branes for example), except in the special case of rigid Calabi-Yau manifolds in IIA~\cite{DeWolfe:2005uu}.  One would like to understand if there are other tools that can be used to construct models with richer structure.

We get a hint about such possibilities by considering T-duality.  If we have a flux compactification which enjoys a circle isometry, then T-duality allows us to find a new vacuum of the theory.  This new solution will be physically equivalent to the first (especially from a four-dimensional effective point of view), but the ten-dimensional description may appear quite different.  For example, if the original compactification was a torus with $H$-flux which was nonvanishing in the direction of the isometry, then in the new solution the $H$-flux will be replaced by a twisting of the circle fibration which is described by non-constant metric components.  This new solution is known as a twisted torus, and the data describing it (analogous to the data describing the $H$-flux) is called metric (or sometimes geometric) flux.  One might then perform a second T-duality introducing the possibility that the resulting compactification is no longer a manifold at all (transition functions between patches are no longer diffeomorphisms, but rather sit inside the T-duality group of string theory), though it is still sensible as a solution of string theory.  Such a situation is often referred to as a nongeometric compactification (though this term can have much broader meaning and is probably not the most appropriate here), and the data describing it is called nongeometric flux.

On a torus, these objects are often referred to by their comopnents $H_{ijk}$, $f^i_{jk}$ and $Q^{ij}_k$ for $H$-flux, metric flux, and nongeometric flux respectively.  Collectively, we will refer to these (in any context) as generalized NSNS fluxes, and they can in fact lead to richer structure and new phenomena in the effective theory~\cite{Kaloper:1999yr,Aldazabal:2006up,Micu:2007rd,D'Auria:2007ay,Ihl:2007ah,Palti:2007pm,Robbins:2007yv} (for a review of the ways these objects appear in string theory, please see~\cite{Wecht:2007wu} and references therein).  One might also wish to contemplate performing T-dualities along all three legs of $H$-flux on the torus, producing an object with components of the form $R^{ijk}$~\cite{Shelton:2005cf}.  It is not clear how to carry out this prescription, however, since T-duality is usually understood to act not on $H$ itself, but rather on a particular triviailzing two-form $B$.  By picking a trivialization, we must break at least one of our three isometries.

Once we have turned on some of these generalized NSNS fluxes, we can turn our attention to RR fluxes in these backgrounds.  String theory includes a set of differential form gauge potentials known as RR fields, whose field strengths can be thought of as elements of the real cohomology of the spacetime manifold.  In fact, upon quantization it has been shown that these field strengths, or fluxes, should actually be thought of as elements of a K-theory group (or some variant thereof for different string theories) of spacetime.  A combination of the A-roof genus and the Chern character maps these K-theory classes into cohomology classes.

In the presence of $H$-flux, it is understood that the correct classification of RR fluxes is given by twisted K-theory, $K_H^i(X)$, where $X$ is spacetime, and where the degree $i$ is $0$ or $1$ for type IIA or IIB respectively.  Upon application of the Chern character map mentioned above this twisted K-theory maps to twisted cohomology in which we have the usual forms on $X$, but the differential $d$ gets replaced by $d_H$, which acts as
\be
d_H\al=d\al-H\w\al.
\ee
The focus of the current work is to understand how this structure can be adapted for the case of generalized NSNS fluxes leading to a modified differential $\mathcal{D}$ encoding the data of the generalized fluxes. Such a differential was first discussed in \cite{Shelton:2006fd}. This paper will be a first step in which we focus primarily on the compactifications $X$ with generalized fluxes and the twisted cohomology theories describing the unquantized RR fluxes.  A sequel paper~\cite{sequel} will deal with the issue of quantization and twisted K-theory.

To get a better sense of what we will be dealing with, let us illustrate how metric fluxes can arise in a simple geometric setup.  Let us assume that our spacetime manifold has the form of a $d$-dimensional torus $T^d$ fibered over a base $X$.  The transition functions of this fibration should be maps from overlaps $U_{ij}$ to the diffeomorphism group of the torus, $\operatorname{Diff}(T^d)$.  However, it can be shown that the group $\operatorname{Diff}(T^d)$ is homotopic to $\SL(d,\Z)\ltimes\U(1)^d$, where the $\SL(d,\Z)$ factor is a diffeomorphism of the lattice, and the $U(1)^d$ factor represents translations on the torus.  Thus, to specify a torus fibration, we can restrict our transition function to be in the smaller group. Let $E$ be the total space of this fibration.  To specify $E$, we should first specify the $\SL(d,\Z)$ elements of each transition function, thus describing an $\SL(d,\Z)$-principal bundle over $X$.  If this bundle is trivial, we next describe a set of integral two-forms $e^a\in H^2(X,\Z)$ giving the Euler class for each circle.  If the principal bundle part is nontrivial, then these combine into a twisted cohomology class discussed below.  Similarly, if we also wish to include $H$-flux in this construction, we must specify an element of $H^3(E,\Z)$.  We will see below how this is equivalent to picking an element of $H^3(X,\Z)$ (modulo a small technical point) and a set of $d$ elements $\om_a\in H^2(X,\Z)$ (again, for the $\SL(d,\Z)$ bundle nontrivial, an element in a twisted cohomology), along with a consistency condition between the $\om_a$ and the $e^a$ (and the $\SL(d,\Z)$ principal bundle).

In string theory, however, our transition functions no longer have to be combinations of diffeomorphisms of the fiber and gauge shifts of the $B$-field~\cite{Hellerman:2002ax}.  Rather, they can include elements in the T-duality group, and our compactifications are hence specified by a $\operatorname{Spin}(d,d,\Z)$ principal bundle over $X$ and a collection of $2d$ elements of $H^2(X,\Z)$ (or again a twisted cohomology class with $2d$ components as we will describe below), along with some consistency conditions.

Indeed, it is useful to consider how these data can be thought of in terms of flux components.  Let's use indices $a$, $b$, to denote fiber indices and indices $\mu$, $\nu$, $\lambda$ to denote base indices.  We can specify some $H$-flux on the base, $H\in H^3(X,\Z)$, with components $H_{\mu\nu\lambda}$.  Our $2d$ elements of $H^2(X,\Z)$ can be thought of as fluxes with two legs on the base and one leg on the fiber, and they can be further split into $d$ contributions to $H$-flux, $H_{\mu\nu a}$, and $d$ metric fluxes $f^a_{\mu\nu}$.  Finally, the principal bundle is classified up to isomorphism by a class in $H^1(X,\operatorname{Spin}(d,d,\Z))$. In the case that this principal bundle is trivial as a $\operatorname{Spin}(d,d)$ principal bundle, we can associate this class with a flat connection in $\Om^1(X,\mathfrak{so}(d,d))$ that can be broken into components corresponding to $H$-flux $H_{\mu ab}$, metric flux $f^a_{\mu b}$, and nongeometric flux $Q^{ab}_\mu$.  Note that this classification never provides any of the $R$-fluxes mentioned above, and in fact does not allow all possible components of the other generalized fluxes either.  We are restricted to components with at least one leg on the base.  This latter restriction does not appear if we only consider the action of twisting the differential, for we can easily write down an operator $\mathcal{D}$ which includes these missing flux components, but it is not clear how this would fit into the framework we discuss.

The goal of this paper and the next is to understand how to construct models involving generalized NSNS fluxes and RR fluxes, and in particular to correctly understand the consistency and quantization conditions demanded by string theory.  Once this foundation has been strengthened, the hope is that models with rich structure in four-dimensions can be understood and constructed and that we can learn more about the space of interesting string theory vacua.

\subsection{Introduction for mathematicians}

It is well-known that the orientation-preserving diffeomorphism group of a $d$-dimensional torus is disconnected and that its group of components is $\mathrm{SL}(d,\BZ)$. Thus, to describe an oriented $T^d$ bundle, we need two pieces of data, a principal $\mathrm{SL}(d,\BZ)$ bundle and a set of classes that describe how the individual circles fiber over the base. The compatibility of these two pieces of data can be described as follows. Let $X$ be a smooth manifold and $E$ an $\mathrm{SL}(d,\BZ)$ principal bundle over it. $\mathrm{SL}(d,\BZ)$ preserves a $\BZ^d$ lattice in $\BR^d$ which we denote $L$. This representation defines a system of local coefficients which we also denote by an abuse of notation, $L$, and we can form the cohomology, $H^\bullet(X,L)$. The twisting of the circles is described by a class in $H^2(X,L)$. If $E$ is a trivial bundle, this reduces to $d$ integral two-forms, the Euler classes for the circle bundles that make up the $T^d$ bundle. The twisted forms describe twistings of the circles compatible with the $\mathrm{SL}(d,\BZ)$ twists.

In string theory, tori have further ``non-geometric" automorphisms. For $T^d$, these are given by the discrete group $\Spin(d,d,\BZ)$. This is called the T-duality group. We can describe a string-theoretic moduli space of tori as the locally symmetric space
\begin{equation}
\M=\SO(d,d;\Z)\backslash\SO(d,d)/S\!\lp\O(d)\times\O(d)\rp\ .
\end{equation}
The goal of this paper is to describe a cohomology theory for these non-geometric torus fibrations.

Let $V = \BR^d$ and $W = V \oplus V^*$. Then $W$ has a natural inner product of signature $(d,d)$. $\Spin(d,d,\BZ)$ acts on $W$ preserving the inner product and an integral lattice, $L$.
 Let $E$ be a principal $\Spin(d,d,\BZ)$ bundle on $X$. As above, we obtain a local system, $L$, and a cohomology
 \begin{equation}
H^\bullet(X,L)\ .
\end{equation}

The inner product on $W$ gives rise to a map:
\begin{equation}
\langle\cdot,\cdot\rangle_2 : H^p(X,L) \otimes H^q(X,L) \to H^{p+q}(X,\BZ)\ .
\end{equation}
(The subscript of 2 is there because this inner product differs by a factor of two from the inner product more natural for the Clifford algebra given in \eqref{cliffprod}.) As the norm on $L$ is even, we can define $||\cdot||^2 : H^p(X,L) \to H^{2p}(X,\BZ)$ by
\begin{equation}
||\sigma||^2 = {\textstyle\frac{1}{2}}\langle\sigma,\sigma\rangle_2\ .
\end{equation}

Next, let $\theta$ be an exact element in $H^k(X,\BZ)$. Let $k_\theta$ be the subspace of $\alpha \in C^{k-1}(X,\BZ)$ such that $d\alpha = \theta$, and let $K_\theta = k_\theta / (\alpha \sim \alpha + d\beta)$ where $\beta \in C^{k-2}(X,\BZ)$. Then $K_\theta$ is an $H^{k-1}(X,\BZ)$ torsor.

\begin{defn}
A \textbf{T-fold} over a manifold, $X$, is given by the following data:
\begin{enumerate}
\item A $\Spin(d,d,\BZ)$ principal bundle, $E$ over $X$,
\item an element $\sigma \in H^2(X,L)$ such that
\begin{equation}
\label{innprodcond}
||\sigma||^2 = [0]\ ,
\end{equation}
\item and an element $\gamma \in K_{||\sigma||^2}\ $.
\end{enumerate}
\end{defn}

Now, let $\mcT$ be a T-fold. In the text, we define a $\BZ_2$-graded cohomology theory $H_\mcT^\bullet(X,\BR)$. It has an obvious automorphism given by the T-duality group $\Spin(d,d,\BZ)$. The physical interpretation of this cohomology theory is that the unquantized Ramond-Ramond fields in string theory live in it. We show how it can be defined either in terms of twisting the differential or in terms of forms on open sets with the usual differential and obeying certain transition functions. In the follow-up to this paper \cite{sequel}, we will use the latter description to define a twisted version of K-theory which will classify the quantized Ramond-Ramond fields.

To show that this cohomology reproduces the correct notion in a geometric situation, recall that $\mathrm{SL}(d,\BZ)$ embeds into $\Spin(d,d,\BZ)$. Suppose we have a T-fold $\mcT$ such that the principal bundle $E$ is associated to a $\mathrm{SL}(d,\BZ)$ principal bundle, $\widetilde{E}$. Then the lattice $L$ decomposes into $\widetilde{L}\oplus \widetilde{L}^* \subset W$ with $\widetilde{L}$ a lattice in $V$, and we have
\begin{equation}
H^\bullet(X,L) \cong H^\bullet(X,\widetilde{L}) \oplus H^\bullet(X,\widetilde{L}^*)\ .
\end{equation}
The local system $\widetilde{L}^*$ is associated to the bundle $\widetilde{E}$ by the dual to the fundamental representation. Thus, $\sigma$ can be decomposed as $r+s$ with $s(r) = [0] \in H^4(X,\BZ)$. The data of the principal bundle $\widetilde{E}$ and the class $r \in H^\bullet(X,\widetilde{L})$ define an oriented torus bundle\footnote{The class $r$ differs by a sign from the Euler class above.} over $X$ which we will denote $T$. The condition $s(r) = [0]$ along with the element $\gamma \in K_{||\sigma||^2}$ tells us that there exists a class $H \in H^3(T,\BZ)$. There exists a natural map $\pi^{\widetilde{L}}_* : H^\bullet(T,\BZ) \to H^\bullet(X,\widetilde{L}^*)$ given by integrating over a particular fiber corresponding to an element in $\widetilde{L}$. The class $H$ satisfies $\pi^{\widetilde{L}}_*H = s$. The ambiguity of the addition of a three-form pulled back from the base corresponds to the class $\gamma$. We have the following theorem:

\begin{thm}
Given the data defined as above, we have the following isomorphism:
\begin{equation}
H^\bullet_\mcT(X,\BR) \cong H^\bullet_H(T,\BR)\ .
\end{equation}
Furthermore, in the case where we can choose $H = 0$, $H^\bullet_\mcT(X,\BZ)$ can be defined over the integers and we have
\begin{equation}
H^\bullet_\mcT(X,\BZ) \cong H^\bullet(T,\BZ)\ .
\end{equation}
\end{thm}

Here, $H_H^\bullet(S,\BR)$ is the $\BZ_2$-graded twisted cohomology defined by the differential
\begin{equation}
d_H = d - H \wdg\ ,
\end{equation}
acting on even and odd differential forms.

\subsection{Plan of the paper}

In the next section we recall some facts about Clifford algebras, spin groups, and orthogonal groups in signature $(d,d)$ that will be helpful in understanding the following sections.  In section~\ref{sec:threelegs} we review how the theory with $H$-flux on $X$ can be thought of as a twisting of cohomology.  In section~\ref{sec:oneleg} we consider $\operatorname{Spin}(d,d,\Z)$ principal bundles over $X$, how they can be classified, and under what circumstances they can be thought of as twistings of cohomology on the Cartesian product $X\times T^d$.  In section~\ref{sec:twolegs} we introduce the two-forms on $X$ which describe the fluxes with two legs on the base $X$, and we learn how they can be thought of as twists.  Section~\ref{sec:alllegs} combines the cases of one leg on $X$ and two legs on $X$ by deriving the conditions on the two forms in the case of a nontrivial $\operatorname{Spin}(d,d,\Z)$ principal bundle.  Finally, section~\ref{sec:discuss} provides a discussion of the results and future directions, including an advertisement of results in the sequel paper.

\section{Clifford algebra actions}\label{sec:clifford}

We will begin by recalling some facts about $\VV$, where $V$ is a real vector space of dimension $d$, largely following~\cite{Gualtieri:2003dx}.  We will later have in mind the case where $V$ is the tangent space to $T^d$ at a point.

\subsection{Linear algebra of $\VV$}\label{subsec:linalg}

We have a nondegenerate symmetric bilinear form on $\VV$,
\be
\label{cliffprod}
\langle X+\xi,Y+\eta\rangle=\hlf\lp\xi(Y)+\eta(X)\rp,
\ee
where $X,Y\in V$ and $\xi,\eta\in V^\ast$.  This inner product has signature $(d,d)$, so the noncompact group of linear transformations which preserve this inner product is the orthogonal group $\O(\VV)\cong\O(d,d)$.

We can also construct a natural pairing between $\w^kV^\ast$ and $\w^kV$ by
\be
\lp v^\ast,u\rp=\det\lp v_i^\ast(u_j)\rp,
\ee
where $v^\ast=v_1^\ast\w\ldots\w v_k^\ast\in\w^kV^\ast$ and $u=u_1\w\ldots\w u_k\in\w^kV$.  Since the highest exterior product of $\VV$ can be decomposed as
\be
\w^{2d}(\VV)\cong\w^dV\otimes\w^dV^\ast,
\ee
we can use the pairing between $\w^dV$ and $\w^dV^\ast$ to give a linear map between the one-dimensional vector spaces $\w^{2d}(\VV)\rr\R$.  As this map is clearly nonzero, it is a canonical isomorphism of vector spaces.  The number $1\in\R$ then defines a distinguished element of $\w^{2d}(\VV)$, and, hence, a canonical orientation on $\VV$.  The group preserving both the symmetric bilinear form and the canonical orientation is the special orthogonal group $\SO(\VV)\cong\SO(d,d)$.

The group $\SO(\VV)$ has two connected components, and the identity component can be obtained by exponentiating the Lie algbra $\mathfrak{so}(\VV)\cong\w^2(\VV)\cong\operatorname{End}(V)\oplus\w^2V^\ast\oplus\w^2V$.  Explicitly, we can write an element $T\in\mathfrak{so}(\VV)$ by its linear action on $\VV$ as
\be
T=\lp\begin{matrix}A & \beta \\ B & -A^\top\end{matrix}\rp,
\ee
where $A\in\operatorname{End}(V)$, $B$ is a two-form in $\w^2V^\ast$ acting by $B(X)=\iota_XB$ for $X\in V$, and $\beta$ is an antisymmetric bivector in $\w^2V$ acting similarly.  We will now see how to exponentiate some simple classes of elements $T$.

Given any $B\in\w^2V^\ast$, let
\be
T_B=\lp\begin{matrix}0 & 0 \\ B & 0\end{matrix}\rp.
\ee
Then
\be
M_B=\exp(T_B)=\lp\begin{matrix}1 & 0 \\ B & 1\end{matrix}\rp\in\SO(\VV),
\ee
\ie $M_B\cdot(X+\xi)=X+(\xi+\iota_XB)$, for $X\in V$, $\xi\in V^\ast$.  We will see below that this family generates $B$-field shifts in string theory on $T^d$.

Similarly,
\be
T_\beta=\lp\begin{matrix}0 & \beta \\ 0 & 0\end{matrix}\rp,\qquad M_\beta=\exp(T_\beta)=\lp\begin{matrix}1 & \beta \\ 0 & 1\end{matrix}\rp,
\ee
so $M_\beta\cdot(X+\xi)=(X+\iota_\xi\beta)+\xi$.  In the string theory context this family has no purely geometric interpretation and mixes the metric and the $B$-field.

Finally, for $A\in\operatorname{End}(V)$,
\be
T_A=\lp\begin{matrix}A & 0 \\ 0 & -A^\top\end{matrix}\rp,\qquad M_A=\exp(T_A)=\lp\begin{matrix}\exp A & 0 \\ 0 & (\exp A^\top)^{-1}\end{matrix}\rp.
\ee
In particular, in this way we can embed any element $P$ of the identity component $\GL^+(V)$ of $\GL(V)$ into the identity component of $\SO(\VV)$,
\be
M_P=\lp\begin{matrix}P & 0 \\ 0 & (P^\top)^{-1}\end{matrix}\rp.
\ee
This mapping can of course be extended to an inclusion of the full $\GL(V)$ group into $\SO(\VV)$.  These will correspond to basis changes in string theory on the torus.

These families of elements are enough to generate all of $\SO(\VV)$.  For example, T-duality of both circles of a string theory two-torus is associated to the $\SO(2,2)$ element
\be
M_T=\lp\begin{matrix}0 & \begin{smallmatrix}1 & 0 \\ 0 & 1\end{smallmatrix} \\ \begin{smallmatrix}1 & 0 \\ 0 & 1\end{smallmatrix} & 0\end{matrix}\rp=\lp\begin{matrix}\s & 0 \\ 0 & \s\end{matrix}\rp\lp\begin{matrix}1 & 0 \\ \s & 1\end{matrix}\rp\lp\begin{matrix}1 & \s \\ 0 & 1\end{matrix}\rp\lp\begin{matrix}1 & 0 \\ \s & 1\end{matrix}\rp,
\ee
where $\s=\lp\begin{smallmatrix}0 & -1 \\ 1 & 0\end{smallmatrix}\rp$.

\subsection{The Clifford algebra}\label{subsec:cliffalg}

Next we would like to define the Clifford algebra $\cl(\VV)$. It is the free algebra generated by elements of $\VV$ with the one relation that
\be
v^2=\langle v,v\rangle,\qquad\forall v\in\VV.
\ee
Given a basis of $v_1\dots v_k \in \VV$, one can take as a basis for $\cl(\VV)$ all antisymmetric products $v_1\w\ldots\w v_k$.  In this basis $\cl(\VV)$ has an integer grading in which an antisymmetric product of $k$ $v_a$'s has degree $k$.  Unfortunately, the Clifford multiplication only respects this grading mod two, so as an algebra $\cl(\VV)$ is $\Z_2$-graded.

The vector space of forms $S=\w^\bullet V^\ast$ has an action of $\VV$,
\be
(X+\xi)\cdot\varphi=\iota_X\varphi+\xi\w\varphi,\qquad X\in V,\quad\xi\in V^\ast,\quad\varphi\in S,
\ee
which can then be extended to an action of $\cl(\VV)$ on $S$ since
\bea
(X+\xi)^2\cdot\varphi &=& \iota_X\lp\iota_X\varphi+\xi\w\varphi\rp+\xi\w\lp\iota_X\varphi+\xi\w\varphi\rp\non\\
&=& \lp\iota_X\xi\rp\varphi\\
&=& \langle X+\xi,X+\xi\rangle\varphi.\non
\eea

The canonical orientation $\nu\in\w^{2d}(\VV)$ also sits inside $\cl(\VV)$ and satisfies $\nu^2=1$, so we can decompose $S$ into positive and negative eigenspaces of $\nu$
\be
S=S^+\oplus S^-,
\ee
and it can be checked that $S^+=\w^{\mathrm{ev}}V^\ast$ and $S^-=\w^{\mathrm{odd}}V^\ast$, \ie this is the decomposition into forms of even or odd degree.

Now we would like to find the group $\SO(\VV)$ inside the Clifford algebra.  First note that the set of multiplicatively invertible elements of the Clifford algebra form a group which is generated (multiplicatively) by $\R^\times$ and those elements of $\VV$ with nonzero norm.  What we really need is a group that can act on $\VV$, so we define the Clifford group $\mathcal{C}$ of elements $x$ such that $xv(x^\ast)^{-1}\in V$, for all $v\in V$, and where\footnote{Of course this only defines $x^\ast$ for elements $x$ of pure even or odd degree, but we can extend the action to all of $\cl(\VV)$ by linearity.  It turns out, however, that the elements of $\mathcal{C}$ with this definition are all of pure degree.} $x^\ast=(-1)^{\operatorname{deg}(x)}x$.  Note that this action of $\mathcal{C}$ on $\VV$ preserves the bilinear form and hence defines a map from $\mathcal{C}$ to $\O(\VV)$ via the homomorphism $\rho:\mathcal{C}\rr\O(\VV)$ given by
\be
\rho(x)\cdot v=xv(x^\ast)^{-1},\qquad x\in\mathcal{C},\quad v\in\VV.
\ee
The kernel of this map is $\R^\times$.  To remove the redundancy of multiplication by $\R^{>0}$, we define the group
\be
\operatorname{Pin}(\VV)=\left\{v_1\cdots v_k|v_i\in\VV,\langle v_i,v_i\rangle=\pm 1\right\},
\ee
which is a double cover of $\O(\VV)$.  The subgroup of $\operatorname{Pin}(\VV)$ consisting of elements satisfying $x=x^\ast$ (which are simply the elements above with $k$ even) is called $\Spin(\VV)$ and can be written
\be
\Spin(\VV)=\left\{v_1\cdots v_{2k}|v_i\in\VV,\langle v_i,v_i\rangle=\pm 1\right\}.
\ee

This group is a double cover of $\SO(\VV)$ via the homomorphism $\rho$ defined above restricted to $\Spin(\VV)\subset\mathcal{C}$, and it preserves the two representations $S^+$ and $S^-$.  The Lie algebra of $\Spin(\VV)$ is $\w^2(\VV)\subset\cl(\VV)$ and acts on $\VV$ by commutation, $x\cdot v=[x,v]$, where $x\in\w^2(\VV)$ and $v\in\VV$.  This action must coincide with the $\mathfrak{so}(\VV)$ actions discussed in the previous subsection, and so we are able to identify the families of elements that we found above.  

To understand their action, let $\eta_a$ be a basis for $V$, and let $\eta^a$ be the corresponding dual basis for $V^\ast$. Recall that given $B=\hlf B_{ab}\eta^a\w \eta^b\in\w^2V^\ast$, we defined $T_B\in\mathfrak{so}(\VV)$ by the action $T_B\cdot(X+\xi)=\iota_XB$, so we can identify
\be
T_B=\hlf B_{ab}\eta^b\eta^a\in\cl(\VV).
\ee
This element acts on a form $\varphi\in S$ by
\be
T_B\cdot\varphi=\hlf B_{ab}\eta^b\w \eta^a\w\varphi=-B\w\varphi.
\ee
We can then exponentiate this action to find
\be
N_B\cdot\varphi=\exp\lp T_B\rp\cdot\varphi=e^{-B}\w\varphi.
\ee
Note that $\rho(N_B)=\rho(-N_B)=M_B\in\SO(\VV)$.

Similarly, for $\beta=\hlf\beta^{ab}\eta_a\w \eta_b$, we had $T_\beta\cdot(X+\xi)=\iota_\xi\beta$, so $T_\beta=\hlf\beta^{ab}\eta_b\eta_a$, and hence
\be
T_\beta\cdot\varphi=\hlf\beta^{ab}\iota_b\iota_a\varphi=\iota_\beta\varphi,
\ee
where $\iota_\beta$ is contraction against the bivector $\beta$.  Exponentiating, we find
\be
N_\beta\cdot\varphi=\exp\lp T_\beta\rp\cdot\varphi=e^{\iota_\beta}\varphi.
\ee

Finally, we consider the $\mathfrak{gl}(V)$ elements.  We had $T_A\cdot(X+\xi)=A(X)-A^\top(\xi)$, so if we write $A=A^b_a\eta^a\otimes \eta_b$ we must have $T_A=\hlf A^b_a(\eta_b\eta^a-\eta^a\eta_b)$,
\be
T_A\cdot\varphi=\hlf A^b_a\lp\iota_b\lp \eta^a\w\varphi\rp-\eta^a\w\iota_b\varphi\rp=\hlf A^a_a\varphi-A^b_a\eta^a\w\iota_b\varphi=-A^\top\varphi+\hlf\lp\Tr A\rp\varphi.
\ee
By exponentiation, we learn that for any element $P\in\GL^+(V)$, we have
\be
\label{pullbackact}
N_P\cdot\varphi=\sqrt{\det P}\lp P^\top\rp^{-1}\varphi,
\ee
where $(P^\top)^{-1}\varphi$ is 
%the pullback by the fundamental representation, \ie 
the usual action of the dual representation on the exterior product $\wdg^\bullet V^\ast$.  This implies that as a $\GL^+(V)$ representation, $S=(\w^\bullet V^\ast)\otimes(\det V)^{1/2}$.

Finally, to get the rest of the group $\Spin(\VV)$, we must find the double cover of the non-identity component of $\SO(\VV)$, and it is sufficient to find the pre-image under $\rho$ of each element $P$ in the nonidentity component $\GL^-(V)$ of $\GL(V)$.  In the identity component, each element $M\in\SO^+(\VV)$ led to a preferred element $N_M\in\rho^{-1}(M)\subset\Spin^+(\VV)$ by exponentiation, and  then $\rho^{-1}(M)=\{N_M,-N_M\}$.  In the non-identity component however, there is no preferred choice, so we must make an arbitrary choice of sign.  We will take the convention
\be
N_P\cdot\varphi=\lp-\det P\rp^{1/2}\lp P^\top\rp^{-1}\varphi,
\ee
and we again have $\rho^{-1}(P)=\{N_P,-N_P\}$.

\subsection{String theory context}\label{subsec:stringtheorycontext}

We would now like to see how some of these considerations are relevant for the study of type II string theory on a torus $T^d$.  A closed string state on such a background carries a set of $2d$ quantum numbers, the winding and momentum of the string on each circle.  The torus itself is defined by a $d$-dimensional lattice $\widetilde L\subset\R^d$, and with slight abuse of notation we can identify this $\R^d$ with the tangent space to the torus at a point, which we will call $V$.  Note that the total tangent bundle of $T^d$ is trivial, so that as bundles over $T^d$, we have simply $T(T^d)\cong V\times T^d$.  Then the momenta lie in the dual lattice $\widetilde{L}^\ast\subset V^\ast$, and the windings lie in the original lattice $\widetilde L$.  The full lattice $L=\widetilde L\oplus\widetilde{L}^\ast\subset\VV$ is called the Narain lattice of the torus compactification.  A nice related construction which provides a geometric setting for this lattice can be found in~\cite{Hull:2004in}.

From this description it is clear that there is an inner product on $L$ inherited from the natural inner product on $\VV$ (though in our conventions this inner product is one half times the usual inner product on $L$), and similarly there is a natural orientation.  The group of linear transformations which preserve these structures and also preserve $L$ is the group $\SO(L)\cong\SO(d,d,\Z)$, a discrete subgroup of the group $SO(\VV)\cong SO(d,d)$.  Following the discussion of the Clifford algebra above, we can also look at the pre-image of $\SO(d,d,\Z)$ under $\rho$, and denote this group as $\Spin(L)\cong\Spin(d,d,\Z)$.

By reducing string theory on this torus, we obtain various bosonic fields.  First there are the NSNS moduli, which describe the metric on the torus and the periods of the two-form $B$-field on two-cycles of the torus.  Together these moduli are parametrized by the coset space
\be
\M=\SO(d,d,\Z)\backslash\SO(d,d)/\operatorname{S}\!\lp\O(d)\times\O(d)\rp.
\ee
The RR fields are slightly more subtle.  Suppose the full ten-dimensional manifold of the string target space is $X\times T^d$, where $X$ is a $(10-d)$-dimensional manifold.  Then the RR potential fields are odd (even) forms in IIA (IIB), that is they are sections of $\Om^\bullet(X\times T^d)$.  With the usual exterior derivative $d$, this gives the usual cochain complex for de Rham cohomology.  

It will actually be much more convenient for us to work with a bundle over $X$ alone, forgetting any dependence on the torus coordinates.  Fortunately, the cochain complex described above is homotopic to one with the desired properties, namely the bundle $\Om^\bullet(X)\otimes S=\Om^\bullet(X,S)$ over $X$, where the tensor product is graded commutative, and, with a slight abuse of notation, $S$ represents both the graded vector space $\w^\bullet V^\ast$ described above and the trivial bundle\footnote{Eventually, we will replace the trivial bundle $S$ by a nontrivial bundle $\mcV$. See section \ref{sec:oneleg}.} of this vector space over $X$.  We still use the usual exterior derivative, but now since nothing depends on the coordinates of the torus, $d$ refers just to the exterior derivative on $X$.  One can view this new complex as a restriction to forms whose Lie derivative vanishes along each of the vectors generating our $d$ circle isometries.  Equivalently, one can think in terms of a Fourier expansion of functions on $T^d$, in which case this construction amounts to restricting to the zero mode.

It is perhaps worthwhile to comment on why we work with $\Om^\bullet(X,S)$ rather than\linebreak $\Om^\bullet(X\times T^d)$.  Following our earlier discussions, it seems natural to take a basis $\eta^a=d\theta^a\w\cdot$ (where $\theta^a$ are coordinates on $T^d$) and $\eta_a=\iota_{\p/\p\theta^a}$ for $V^\ast$ and $V$ in $\cl(\VV)$.  Unfortunately, if $d$ is the full differential on $X\times T^d$, then $\{d,\eta^a\}=0$, but $\{d,\eta_a\}=\mathcal{L}_{\p/\p\theta^a}$ does not vanish in general.  This will be problematic since we will want to build twisted differentials $\mathcal{D}=d+\mathcal{A}$, where $\mathcal{A}$ involves Clifford algebra operations.  We will require $\mathcal{D}^2=0$, but as it stands, $\mathcal{D}^2$ may contain terms proportional to Lie derivatives, arising from $\{d,\mathcal{A}\}$.  An alternative is to take $\eta_a\al=-(-1)^{\mathrm{deg}\,\al}\pi_{a\,\ast}\al$, where for $\al=\al_0+\al_1\w d\theta^a$,
\be
\pi_{a\,\ast}\al=\frac{1}{2\pi}\int_0^{2\pi}\al_1d\theta^a,
\ee
but this has the disadvantage that $\{\eta^a,\eta_b\}=\delta^a_b\operatorname{Av}_a$, where
\be
\operatorname{Av}_a(\al)=\frac{1}{2\pi}\lp\int_0^{2\pi}\al_0d\theta^a\rp+\frac{1}{2\pi}\lp\int_0^{2\pi}\al_1d\theta^a\rp\w d\theta^a,
\ee
is an averaging operator.  In other words, the Clifford algebra does not have the right anticommutation relations.   Since the averaging operator is homotopic to the identity, this can be thought of as a homotopy action of the Clifford algebra on the dga $\Omega^\bullet(T^d)$. This leads to many technical complications, however, so we invert the homotopy by hand and work with forms that have no dependence on $\theta^a$. With this condition, both problems vanish; $\mathcal{L}_a=0$, $\operatorname{Av}_a=1$, and the two definitions of $\eta_a$ are equivalent.

\section{Three legs on the base}\label{sec:threelegs}

As a warm up for the more complicated nongeometric situation, let us begin with the simple case of a three form NSNS flux. This is specified by
\begin{equation}
[H] \in H^3(X,\BZ)\ .
\end{equation}
As is well known, with this flux turned on, the supergravity equations for the gauge-invariant Ramond-Ramond fields imply that
\begin{equation}
dF_\mathrm{tot} = H \wdg F_\mathrm{tot}
\end{equation}
where $F_\mathrm{tot}$ is the total Ramond-Ramond flux, and $H$ is a three form representing the de Rham cohomology class of $[H]$. This motivates one to consider the following cohomology theory. Let $\Omega^e(X)$ ($\Omega^o(X)$) be the spaces of differential forms on $X$ of even (odd) degree. Given a cochain $H \in C^3(X,\BZ)$, $dH = 0$, we can define the following ``twisted" differential:
\begin{equation}
d_H\alpha = d\alpha - H \wdg \alpha
\end{equation}
where we have identified $H$ with a closed differential form. It is straightforward to determine that this squares to zero (note that it does \textit{not} square to zero over the integers) and thus gives rise to a cohomology theory which we will denote $H^\bullet_H(X,\BR)$. This cohomology theory was first studied in the physics literature in \cite{Rohm:1985jv}. By its very (rational) nature, it cannot serve as a proper repository for quantized RR-fields. Instead, RR-fields live in twisted K-theory, and there exists a Chern character that maps to this twisted cohomology.

Before proceeding, notice that the twisted differential $d_H$ depends on an actual representative of $H$ in the cohomology class $[H]$. If we have a different differential given by $d_{H'}$ with $H' = H + dB$, then we have a canonical isomorphism $H_H^\bullet(X,\BR) \cong H_{H'}^\bullet(X,\BR)$ given by $\alpha \mapsto e^{-B}\alpha$. Thus, we can talk about the twisted cohomology group with respect a class $[H] \in H^3(X,\BZ)$.

In this section, we will describe a local description of this cohomology theory due to Atiyah and Segal \cite{Atiyah:2005gu} in terms of open sets and transition functions. Let the collection of sets $U_i$ be an open cover of $X$. We will denote by $U_{i_1\dots i_k}$ the intersection $U_{i_1} \cap \dots \cap U_{i_k}$. Furthermore, we will require that this cover is a ``good cover" in the sense that the $U_i$ and all their intersections are contractible sets. (This is stronger than we need, in fact, but every manifold has a good cover \cite{BT}.) Since each $U_i$ has no cohomology (the Poincar\'{e} lemma), the restriction of $H$ to each $U_i$ is exact. Thus, we have
\begin{equation}
H_i = dB_i\ .
\end{equation}
On each overlap $U_{ij}$ we can define $B_{ij} = B_i - B_j$. Since $H$ is a globally defined form, we have that $dB_{ij} = 0$. In addition, we have $B_{ij} = - B_{ji}$ and the cocycle condition, $B_{ij}  + B_{jk} + B_{ki} = 0$.

We can now define the following cohomology theory. A cochain will be a collection of forms $\alpha_i \in \Omega^\bullet(U_i)$ which obey the following transition function on the overlaps, $U_{ij}$:
\begin{equation}
\alpha_j = e^{B_{ij}} \alpha_i\ .
\end{equation}
The differential will be the application of the usual de Rham differential on each open set. Because $B_{ij}$ is closed, this respects the transition functions. This cohomology theory is, in fact, isomorphic to $H^\bullet_H(X,\BR)$. This can be seen as follows. Given a set of forms $\alpha_i$ as above, on each $U_i$, define 
\begin{equation}
\label{hatmap1}
\widehat{\alpha}_i = e^{B_i} \alpha_i\ .
\end{equation}
 Then, on the overlap $U_{ij}$, we have
\begin{equation}
\widehat{\alpha_j} = e^{B_j} \alpha_j = e^{B_j} e^{B_i - B_j} \alpha_i = e^{B_i} \alpha_i = \widehat{\alpha_i}\ ,
\end{equation}
so this is a globally defined form. Furthermore, it obeys
\begin{equation}
d\widehat{\alpha}_i = d\left(e^{B_i} \alpha_i \right) = dB_i \wdg e^{B_i} \alpha_i + e^{B_i} d\alpha_i= H_i \wdg \widehat{\alpha_i} + e^{B_i} d\alpha_i\ .
\end{equation}
We can rewrite this as $d_H\widehat{\alpha} = \widehat{d\alpha}$, so we have a map of complexes.
As the map \eqref{hatmap1} is invertible, this gives rise to an isomorphism on cohomology.

In fact, all we needed for the previous construction was the collection of transition functions $B_{ij}$. Let $\lambda_i$ be a partition of unity subordinate to the open cover $U_i$. In other words, $\lambda_i$ is a collection of functions on $X$ such that $\lambda_i = 0$ on the complement of $U_i$ and $\sum_i \lambda_i = 1$. Then, given the collection $B_{ij}$, we can define
\begin{equation}
B_i = \sum_j \lambda_i B_{ij}\ .
\end{equation}
We can compute
\begin{equation}
B_i - B_j = \sum_k \lambda_k B_{ik} - \sum_k \lambda_k B_{jk} = \sum_k \lambda_k (B_{ik} + B_{kj})\ .
\end{equation}
Because the $B_{ij}$ are transition functions, they obey the cocycle condition $B_{ij} + B_{jk} + B_{ki} = 0$. Thus, we have
\begin{equation}
B_i - B_j = \sum_k \lambda_k B_{ij} = B_{ij}\ ,
\end{equation}
and these provide candidate $B_i$ for the procedure in the previous paragraph. Finally, let $H_i = dB_i$. Then,
\begin{equation}
H_i - H_j = \sum_k d\lambda_k B_{ik} - \sum_k d\lambda_k B_{jk} = d\left(\sum_k \lambda_k\right) B_{ij} = 0\ ,
\end{equation}
so we have a global form $H$.

%The collection $B_{ij}$ should remind the reader of transition functions of a fiber bundle, and in fact 
%they are. The twisted cohomology theory is related to a fibering of a set of spaces called the spectrum
% of the cohomology theory over the base manifold $X$. This perspective will be investigated in the 
%follow-up to this paper, \cite{BRtoap}. In the remainder of this paper, we will show how the cohomology
% theories for non-geometric fluxes can be written in terms of transition functions for forms on an open
% cover. In addition to elucidating various properties of the theory, it will allow us in \cite{BRtoap} to
% define K-theory in these non-geometric backrounds.

\section{One leg on the base}\label{sec:oneleg}

Now we move to the constructions of primary interest, in which a $d$-dimensional torus is fibered over a base space $X$.  In this section we consider the case when the torus bundle is flat, but its moduli (the metric and $B$-field) are allowed to vary over the base.  Our philosophy follows closely that of~\cite{Hellerman:2002ax,Dabholkar:2002sy,Flournoy:2004vn,Hull:2004in,Ihl:2007ah}.  Consider going around a loop in the base $X$.  The $T^d$ fiber should come back to a physically equivalent configuration, \ie the moduli should be related by an $\SO(d,d,\Z)$ transformation.  Lifting to an action also on the RR fields, we need a $\Spin(d,d,\Z)$ transformation.  Clearly (because the groups are discrete) homotopic loops must induce the same transformation, so for every element of $\pi_1(X)$ we should get an element of $\Spin(d,d,\Z)$.  Also, making a global duality transformation (applying the same $\Spin(d,d,\Z)$ transformations to all fibers simultaneously) will give a physically equivalent compactification, so we can argue that, topologically, these torus bundles are classified by
\be
\operatorname{Hom}\lp\pi_1(X),\Spin(d,d,\Z)\rp/\Spin(d,d,\Z),
\ee
where the quotient action is conjugation.

But in fact this is a common situation, and we have, for a general group $G$, the bijections
\be
\operatorname{Hom}\lp\pi_1(X),G\rp/G\cong H^1(X,G)\cong \operatorname{FBun}(X,G),
\ee
where $H^1$ here can be defined, for example, by a \v{C}ech description with locally constant cocycles, and where $\operatorname{FBun}(X,G)$ is the set of isomorphism classes of flat principal $G$-bundles over X.  Flat in this context simply refers to the fact that the transition functions are locally constant.  If $G$ is discrete, as in our case, then the transition functions are necessarily locally constant; all principal bundles are then flat.

For the rest of this discussion we will use $\G$ to denote our duality group, $\Spin(d,d,\Z)$, and we will use $G$ to denote the continuous group $\Spin(d,d)$.  The discussion below actually does carry through to the case of more general examples of discrete $\G\subset G$.  Indeed, this should be the language needed to describe other situations of interest in string theory, including compactifications with K3 fibers~\cite{Cvetic:2007ju}, heterotic compactifications with toroidal fibers, or manifolds patched together by U-duality transformations~\cite{Kumar:1996zx,Hull:2007zu} (equivalently M-theory compactifications with toroidal fibers, perhaps along the lines of~\cite{Bergman:2004ne}).

We have seen that our flat torus bundles are classified by principal $\G$-bundles over $X$.  Let $E$ be such a bundle.  Take a nice open cover of $X$ by sets $U_i$, and let the transition functions of $E$ be $R_{ij}\in\G$ on each overlap $U_{ij}=U_i\cap U_j$.  Since $\G$ is discrete, these transition functions are locally constant, \ie constant maps from $U_{ij}\rr\G$.  Now since $\G\subset G$, we can also view these as transition functions for a principal $G$-bundle $Q$ over $X$, namely the associated $G$ bundle to $E$ given by
\be
Q=E\times_\G G=E\times G/\{(e,g)\sim(e\cdot\g,\g^{-1}h),\forall\g\in\G\}.
\ee
The projection $\pi^Q:Q\rr X$ is given in terms of the projection operator $\pi^E$ for $E$ by $\pi^Q([(e,g)])=\pi^E(e)$.  Note that we have a natural inclusion $i:E\hookrightarrow Q$ given by $i(e)=[(e,1)]$.  The right $G$-action is simply $[(e,g)]\cdot h=[(e,gh)]$.

We can also construct a natural connection on $Q$ as follows.  For any point $q\in Q$ we can pick a representative $(e,g)\in E\times G$, so that we have $q=i(e)\cdot g$.  Then we can define the horizontal subspace of $T_qQ$ by
\be
H_qQ=R_{g\,\ast}i_\ast T_eE,
\ee
where $R_g$ is right multiplication by $g$.  Note that this does not depend on the choice of representative since
\be
R_{\g^{-1}g\,\ast}i_\ast T_{e\cdot \g}E=\lp R_{g\,\ast}R_{\g^{-1}\,\ast}\rp i_\ast\lp R_{\g\,\ast}T_eE\rp=R_{g\,\ast}\lp R_{\g^{-1}}iR_\g\rp_\ast T_eE,
\ee
but
\be
\lp R_{\g^{-1}}iR_\g\rp(e)=\lp R_{\g^{-1}}i\rp(e\cdot \g)=R_{\g^{-1}}([(e\cdot \g,1)])=[(e\cdot \g,\g^{-1})]=[(e,1)]=i(e).
\ee
It is not difficult to check that this does in fact define a (principal Ehresmann) connection on $Q$.  Equivalently, it defines a global one-form $\rho\in\Om^1(Q,\mathfrak{g})$, where $\mathfrak{g}$ is the Lie algbra of $G$.  Indeed, in local coordinates $(e,g)$ around $q\in Q$, we have simply that 
\be
\label{eq:trivialconnection}
\rho=g^{-1}dg.
\ee
Since the transition functions $R_{ij}$ are locally constant, this is a globally defined form.

We will also be interested in other bundles associated to $E$, the ones which describe our moduli.  The NSNS moduli can be thought of as sections of a $G/K$ bundle over $X$ associated to $E$, where\footnote{It is familiar in string theory that the metric and $B$-field moduli of a torus $T^d$ can be parametrized by the coset space $\SO(d,d,\Z)\backslash\SO(d,d)/\operatorname{S}(\O(d)\times\O(d))$.  One can check that this coset is actually isomorphic to the coset $\G\backslash G/K$ described above in terms of the double covers, $\G=\Spin(d,d,\Z)$, $G=\Spin(d,d)$, and the maximal compact subgroup of $G$ is $K=\operatorname{S}(\operatorname{Pin}_+(d)\times_{\Z_2}\operatorname{Pin}_-(d))$.  In this descriptions of $K$, $\operatorname{Pin}_+(d)=\operatorname{Pin}(d,0)$ and $\operatorname{Pin}_-(d)=\operatorname{Pin}(0,d)$ are subgroups of $\operatorname{Pin}(d,d)$ described above.  Their intersection is the subgroup $\Z_2=\{\pm 1\}\subset\cl(d,d)$, and so in the expression for $K$,
\begin{equation*}
\operatorname{Pin}_+(d)\times_{\Z_2}\operatorname{Pin}_-(d)=\operatorname{Pin}_+(d)\times\operatorname{Pin}_-(d)/\{(g,h)\sim(-g,-h)\}.
\end{equation*}
Finally, going to $\operatorname{S}(\operatorname{Pin}(d)\times_{\Z_2}\operatorname{Pin}_-(d))$ just involves restricting to the even degree part of $\cl^\bullet(d,d)$.} $K$ is the maximal compact subgroup of $G$.  In other words, this is the bundle $P=E\times_\G G/K$, or equivalently, $P=Q\times_G G/K$.

For the RR moduli, we will restrict to those forms with no dependence on the torus coordinates, \ie those killed by the Lie derivative along each of the $d$ vectors generating the circle actions.  In this case, as described in section~\ref{subsec:stringtheorycontext}, the RR fields can be viewed as sections of bundles over $X$, whose fiber over each point $x$ is isomorphic to $\w^\bullet T^\ast_xX\otimes S$, where $S=\w^\bullet V^\ast$ is the spinor bundle of $\Spin(d,d)$.  This bundle should of course be twisted globally, so that the forms are sections of $\Omega^\bullet(X,\mathcal{V})$ with
\be
\mathcal{V}=E\times_\G S,
\ee
and $\G$ acts on $S$ as described in section~\ref{subsec:cliffalg}.  Note that we can equivalently write $\mathcal{V}=Q\times_G S$, and that in this way a flat connection on $Q$ gives rise to an associated covariant exterior derivative on $\mathcal{V}$. If $R_{ij} : U_{ij} \to \G$ are a collection of transition functions describing the bundle $E$, a section of $\Omega^\bullet(X,\mathcal{V})$ is given by a collection of forms $\alpha_i \in \Omega(U_i,S)$ such that $\alpha_j = R_{ij}\alpha_i$. The flat connection acts as the usual exterior derivative on these forms. The RR fields then live in the cohomology of this flat connection, which we denote $H^\bullet_\rho(X,\mathcal{V})$.

For general bundles $E$, this seems to be the best we can do.  However, situations can arise in which the principal $G$-bundle, $Q$, associated to $E$ is actually trivial as a principal bundle, even if $E$ is not trivial.  In this case, if we pick a trivializing section $\s$ of $Q$ which induces a global trivialization $\Phi:X\times G\rr Q$ via $\Phi(x,g)=\s(x)\cdot g$, then $\rho'=\Phi^\ast\rho$ is a connection on the trivial principal bundle $X\times G$. The section $\s$ also allows us to construct an isomorphism between $\mathcal{V}$ and the trivial bundle $X \times S$, and the new covariant exterior derivative will be given by
\be
\mathcal{D}(\al\otimes\beta)=d\al\otimes\beta+A\w\al\otimes\beta,
\ee
where $\al\in\Om^\bullet(X)$, and $\beta\in S$ and $A=\s^\ast\rho\in\Om^1(X,\mathfrak{g}_Q)$ is the connection one-form in the base, valued in the adjoint bundle $\mathfrak{g}_Q$ over $X$.  $A$ in this expression acts on $\beta$ via the $\mathfrak{g}$ action on $S$ described in section~\ref{subsec:cliffalg}.  Recalling the representation of $\mathfrak{g}$ in $\cl(\VV)$, we can actually view $A$ as an element of $\Om^1(X,\w^2(\VV))$.  Also, $A$ is covariantly closed, in the sense that $dA+A\w A=0$, and so we can claim that $A$ is the analog of $H$-flux in the case with only one leg on the base $X$.  Indeed, if $A$ actually sits inside the smaller space $\Om^1(X,\w^2V^\ast)$, then we have $dA=0$ and we really can interpret $A$ as $H$-flux on the space $X\times T^d$.

We can see how this works in  a couple of examples.  Suppose our base is a circle, $X=S^1$.  Since $\pi_1(S^1)=\Z$, the principal $\G$-bundles over $X$ are classified by
\be
\operatorname{Hom}(\Z,\G)/\G=\G/\G,
\ee
the set of conjugacy classes of $\G=\Spin(d,d,\Z)$.  A flat principal bundle over $S^1$ is classified by (the conjugacy class of) the monodromy obtained by going once around the circle.  Let $\g\in\G$ be this monodromy.  We would like to know when the associated $G$-bundle can be trivialized.  Let us represent our principal bundle as the space
\be
Q_\g=[0,1]\times G/\{(1,g)\sim(0,g\g^{-1})\}.
\ee
A trivialization of $Q_\g$ is equivalent to a global section $\s:S^1\rr Q_\g$, or equivalently in our setup, a continuous map $\s:[0,1]\rr G$ such that $\s(1)=\s(0)\cdot\g$.  It is clear that if $\g$ does not lie in the identity component of $G$, then this problem has no solution; in this case the bundle is nontrivial.  However, if $\g$ does lie in the identity component of $G$, then we can write $\g=\exp(M)$, for some $M\in\mathfrak{g}=\mathfrak{so}(d,d)$ (not necessarily unique), and in that case we can find a global section,
\be
\s_M(x)=\exp(xM),\qquad x\in[0,1].
\ee
This provides a trivialization, $\Phi_M:X\times G\rr Q_\g$, $\Phi(x,g)=[(x,\exp(xM)\cdot g)]$.  Pulling back our connection, we find
\be
\rho_M'=\Phi_M^*\rho=g^{-1}dg+g^{-1}Mg\,dx=g^{-1}\lp d+M\,dx\rp g,
\ee
and on the base,
\be
A_M=\s_M^\ast\rho=M\,dx.
\ee
Then we can construct a covariant derivative on sections of $\Om^\bullet(X,S)$,
\be
\mathcal{D}_M\lp\al(x)\otimes\beta\rp=d\al\otimes\beta+\lp dx\w\al\rp\otimes\lp M\cdot\beta\rp,
\ee
where $\al(x)\in\Om^\bullet(X)$ and $\beta\in S$, and the action of $M$ on $\beta$ is as described in section~\ref{subsec:linalg}.

Now consider a base $X=T^n$.  In this case we will illustrate a bottom-up construction~\cite{Ihl:2007ah} which generalizes the situation above.  Let $M_i$, $i=1,\ldots,n$ be a set of mutually commuting elements of $\mathfrak{g}$ that in addition satisfy the quantization conditions $\exp(M_i)\in\G$ for each $i$.  Let our $\G$-bundle $E$ be associated to (the equivalence class of) the holonomy map\linebreak $\phi_M:\pi_1(T^n)=\Z^n\rr\G$ given by $\phi_M(k^1,\ldots,k^n)=\exp(k^iM_i)$.  Then we can again construct a global section of $Q$,
\be
\s_M(x^1,\ldots,x^n)=\exp\lp x^iM_i\rp,
\ee
where the $x^i$ parametrize $[0,1]^n$ which we take to cover $T^n$.  As above, this leads to connections on the trivial $G$-bundle and on the base,
\be
\rho'=g^{-1}\lp d+M_i\,dx^i\rp g,\qquad A=M_i\,dx^i,
\ee
so that
\be
\mathcal{D}_M\lp\al(x)\otimes\beta\rp=d\al\otimes\beta+\lp dx^i\w\al\rp\otimes\lp M_i\cdot\beta\rp.
\ee

In fact, if we pick a basis $\{\eta_a,\eta^a\}$ for $\VV$, then we can decompose $M_i$ as
\be
M_i=\lp\begin{matrix}f^a_{ib} & Q^{ab}_i \\ H_{iab} & -f^b_{ia}\end{matrix}\rp,
\ee
and the corresponding differential can be written explicitly as
\bea
\mathcal{D}\lp\al\otimes\beta\rp &=& d\al\otimes\beta-\hlf H_{iab}dx^i\w\al\otimes(\eta^a\cdot\eta^b\cdot\beta)+\hlf Q^{ab}_idx^i\w\al\otimes\lp\eta_b\cdot\eta_a\cdot\beta\rp\non\\
&& +\hlf f^a_{ia}dx^i\w\al\otimes\beta-f^b_{ia}dx^i\w\al\otimes\lp\eta^a\cdot\eta_b\cdot\beta\rp.
\eea
This agrees with with results already in the literature~\cite{Shelton:2006fd,Aldazabal:2006up,Micu:2007rd,Ihl:2007ah}.  Also, the conditions on the fluxes obtained by requiring that $\mathcal{D}^2=0$, equivalently that the connection is flat, reproduce the Bianchi identities from the literature (see the computation of $\mathcal{D}^2$ in~\cite{Shelton:2006fd,Ihl:2007ah}).

\section{Two legs on the base}\label{sec:twolegs}

In this section, we will examine forms with two legs on the base. In particular, we impose that the $Spin(d,d,\BZ)$ bundle, $E$, is trivial. We will see that this situation can be interpreted geometrically.

To be precise, the forms we consider are sections of $\Omega^2(X) \otimes \cl^1(V \oplus V^*)$. However, we need a quantization condition. Recall that we have selected a $\Spin(d,d,\BZ)$ subgroup of $\Spin(d,d)$. This subgroup preserves a lattice in $V \oplus V^*$ which we denote $L$ and can consider as a lattice in $\cl^1$. Then, the quantized forms live in $H^2(X,L)$. The natural inner product on $V \oplus V^*$ (equal to twice the inner product in \eqref{cliffprod}) gives rise to an inner product on forms:
\begin{equation}
\langle\ ,\ \rangle_2 : H^2(X,L) \otimes H^2(X,L) \to H^4(X,\BZ)\ ,
\end{equation}
and an associated norm $||\cdot||^2 : H^2(X,L) \to H^4(X,\BZ)$ given by
\begin{equation}
||\sigma||^2 = {\textstyle\frac{1}{2}}\langle\sigma,\sigma\rangle_2= \langle\sigma,\sigma\rangle\ .
\end{equation}

Given a form $\sigma \in H^2(X,L)$, we require that it obey
\begin{equation}
\label{innprodzero}
||\sigma||^2 = [0]\ .
\end{equation}

If we choose a basis of $V$, $\eta_a$, we obtain a dual basis of $V^*$, ${\eta}^a$. We can expand $\sigma$ in this basis as
\begin{equation}
\label{basisdecomp}
\sigma = \sum_a \lp-e^a \eta_a + \omega_a \eta^a\rp,
\end{equation}
with $e^a, \omega_a \in H^2(X,\BZ)$. In this basis, the condition \eqref{innprodzero} becomes $\sum_a e^a \cup \omega_a = [0]$. The extra sign in this expression is conventional and is related to the definition of the transition functions $s_{ij}$ in what follows.

\subsection{The one dimensional case}

For simplicitly, let us first address the case when $V$ is one dimensional. Then we (canonically) have two two-forms $e,\omega \in H^2(X,\BZ)$. Let us choose a circle bundle, $E$, over $X$ with Euler class $e$ and projection $\pi$. Then, we have the Gysin sequence in cohomology:
\begin{equation}
\cdots \longrightarrow H^3(X,\BZ) {\buildrel \pi^* \over\longrightarrow} H^3(E,\BZ) {\buildrel \pi_* \over\longrightarrow}
H^2(X,\BZ) {\buildrel \cup e \over\longrightarrow} H^4(X,\BZ) \longrightarrow\cdots\ .
\end{equation}
The condition $e \cup \omega = [0]$ implies that there exists a degree three cohomology class $H$ on the total space such that $\pi_*H = \omega$. Thus, we see that we have recovered the geometric situation with three form flux.

Next we would like to see that the twisted (de Rham) cohomology in this situation reproduces the cohomology of the geometric interpretation. If $\omega \wdg e = 0$ as forms (rather than in cohomology), the twisted differential is given by
\begin{equation}
\mcD\alpha = d\alpha - (-)^\alpha e \wdg \pi_*\alpha - \omega \wdg d\theta \wdg \alpha\ .
\end{equation}
As before, $\alpha$ is a form on $X \times S^1$ whose Lie derivative in the fiber direction vanishes, and we have implicitly pulled back $e$, $\omega$ and the push-forward by the standard projection. We will see later how to modify the differential for when the product is zero only in cohomology.

Let us first consider the case where $\omega = 0$ so as to make the geometry more obvious. We again consider a good open cover $U_i$ and trivialize $e$ on each open set: $e_i = df_i$. We can form $f_{ij} = f_i - f_j$. The $f_{ij}$ are closed one forms on the overlaps $U_{ij}$. They can again be trivialized such that $f_{ij} = dg_{ij}$ where $g_{ij}$ is a map from $U_{ij} \to S^1=\R/2\pi\Z$. Define a function $s_{ij} : U_{ij} \times S^1 \to U_{ij} \times S^1$ by $s_{ij}(x,\theta) = (x,\theta - g_{ij}(x))$. These are the transition functions of the circle bundle $E$ defined by the form $e$.

A global form on $E$ can be written as a collection of local forms $\alpha_i \in \Omega^\bullet(U_i \times S^1)$ that are constant in the $S^1$ direction and such that $\alpha_j = s^*_{ij} \alpha_i$. We can write $\alpha$ in local coordinates (and somewhat schematically) as
\begin{equation}
\alpha = a(x) dx_{\mathrm{base}} + b(x) d\theta \wdg dx_{\mathrm{base}}
\end{equation}
Note that $a$ and $b$ do not depend on $\theta$ because of the condition on the Lie derivative.
Then
\begin{equation}
\label{firstFdef}
s_{ij}^*\alpha = \alpha + (-)^\alpha f_{ij} \wdg \pi_*\alpha = e^{-F_{ij}} \alpha
\end{equation}
where $F_{ij}$ is an operator such that $F_{ij}\alpha = -(-)^\alpha f_{ij} \wdg \pi_*\alpha$. This is precisely the operator one obtains by considering $f_{ij}$ as a Clifford algebra valued form. 

We can now use the same prescription as in previous sections to construct a global form from these local forms. In particular, let
\begin{equation}
\label{hatmap2}
\widehat{\alpha}_i = e^{-F_i} \alpha_i
\end{equation}
where $F_i\alpha = -(-)^\alpha f_i \wdg \pi_*\alpha$. Then we have
\begin{equation}
\widehat{\alpha}_j = e^{-F_j}\alpha_j = e^{-F_j} e^{-F_{ij}} \alpha_i = \widehat{\alpha}_i\ ,
\end{equation}
so we have a global form. Its differential is given by
\begin{equation}
d\widehat{\alpha} = d\left(e^{-F_i}\alpha\right) = -(dF_i) e^{-F_i} \alpha + e^{-F_i} d\alpha\ .
\end{equation}
Here, $(dF_i)\alpha = -(-)^\alpha e_i \wdg \pi_*\alpha$. In terms of a twisted differential, this is
\begin{equation}
\mcD\widehat{\alpha} = d\widehat{\alpha} - (-)^\alpha e \wdg \pi_*{\widehat{\alpha}} = \widehat{d\alpha}\ ,
\end{equation}
and we have a map of complexes. The map \eqref{hatmap2} is invertible, so it gives rise to an isomoprhism in cohomology. Thus, we have shown that
\begin{equation}
H^\bullet_\mcD(X,S) \cong H^\bullet(E)\ .
\end{equation}
Recall that our notation for the left hand side refers to taking forms on $X\times S^1$ whose Lie derivative along the circle action vanishes. In fact, in this situation, everything can be defined over the integers, and the proof goes through (this correctly reproduces, for example, the computation of the cohomology of a particular twisted torus in~\cite{Marchesano:2006ns}).

Now, we can include the three form flux corresponding to the form $\omega$. Let us write $e \wdg \omega = d\chi$ and let $\mcA$ be a connection on the total space of $E$. Then,
\begin{equation}
\label{threeformtot}
H = \mcA \wdg \omega - \chi
\end{equation}
is a closed form on $E$ (recall that $d\mcA = e$). On our open cover $U_i$, we can choose $\omega_i = d\lambda_i$ and $\lambda_{ij} = \lambda_i - \lambda_j$. Furthermore, in local coordinates, we can write
\begin{equation}
H_i = (d\theta + A_i) \wdg \omega_i - \chi_i = dB_i
\end{equation}
Note that we can choose $A_i = f_i$, although we have to be careful when comparing on overlaps due to the transformation properties of connections.

We will first find a local model for the $H$-twisted cohomology using the open cover $U_i \times S^1$ of $E$. Our cover is no longer good, but this does not lead to problems. Since we are on a nontrivial bundle, we define $B_{ij} = s^*_{ij}B_i - B_j$. To implement the procedure of section \ref{sec:threelegs}, we define a collection of forms on the $U_i \times S^1$ such that 
\begin{equation}
\label{buntrans}
\alpha_j = e^{B_{ij}} s^*_{ij}\alpha_i
\end{equation}
and define a cohomology theory by taking the de Rham differential restricted to each open set. Then, we can define the form
\begin{equation}
\widehat{\alpha}_i = e^{B_i} \alpha_i
\end{equation}
which is global on $E$ as
\begin{equation}
\widehat{\alpha}_j = e^{B_j} \alpha_j = e^{B_j} e^{B_{ij}} s^*_{ij}\alpha_i = s^*_{ij}\widehat{\alpha_i}\ .
\end{equation}
Furthermore.
\begin{equation}
d_H \widehat{\alpha}_i = \widehat{d\alpha_i}\ ,
\end{equation}
and everything is as before.

To obtain a global form on $X \times S^1$ (as opposed to $E$), we instead define
\begin{equation}
\label{hatmap3}
\overset{\frown}{\alpha}_i = e^{f_i \wdg \lambda_i} e^{B_i} e^{-F_i} \alpha_i\ .
\end{equation}
Then,
\begin{equation}
\begin{split}
\overset{\frown}{\alpha}_j &= e^{f_j \wdg \lambda_j} e^{B_j} e^{-F_j}
\alpha_j
      = e^{f_j \wdg \lambda_j} e^{B_j} e^{-F_j} e^{B_{ij}} s^*_{ij}\alpha_i
\\
      &= e^{f_j \wdg \lambda_j} e^{B_j} e^{-F_j} e^{B_{ij}} e^{-F_{ij}}
\alpha_i
      = e^{f_j \wdg \lambda_j + f_j \wdg \lambda_{ij}} e^{s^*_{ij}B_i}
e^{-F_i} \alpha_i \\
      &= e^{f_j \wdg \lambda_j + f_j \wdg \lambda_{ij} + f_{ij} \wdg
\lambda_i} e^{B_i}e^{-F_i} \alpha_i
      = \overset{\frown}{\alpha}_i\ .
\end{split}
\end{equation}
%\begin{equation}
%\begin{split}
%\overset{\frown}{\alpha}_j &= e^{f_j \wdg \lambda_j} e^{B_j} e^{-F_j} \alpha_j \\
%       &= e^{f_j \wdg \lambda_j} e^{B_j} e^{-F_j} e^{B_{ij}} s^*_{ij}\alpha_i \\
%       &= e^{f_j \wdg \lambda_j} e^{B_j} e^{-F_j} e^{B_{ij}} e^{-F_{ij}} \alpha_i \\
%       &= e^{f_j \wdg \lambda_j + f_j \wdg \lambda_{ij}} e^{s^*_{ij}B_i} e^{-F_i} \alpha_i \\
%       &= e^{f_j \wdg \lambda_j + f_j \wdg \lambda_{ij} + f_{ij} \wdg \lambda_i} e^{B_i}e^{-F_i} \alpha_i \\
%       &= \overset{\frown}{\alpha}_i\ .
%\end{split}
%\end{equation}
Taking the differential, we see that
\begin{equation}
\begin{split}
d\overset{\frown}{\alpha}_i &= d(f_i \wdg \lambda_i) \wdg \overset{\frown}{\alpha}_i + H_i \wdg \overset{\frown}{\alpha}_i -
e^{f_i \wdg \lambda_i} e^{B_i} (dF_i) e^{-F_i} \alpha_i  + \overset{\frown}{d\alpha}_i\\
     &= d(f_i \wdg \lambda_i) \wdg \overset{\frown}{\alpha}_i  + H_i \wdg \overset{\frown}{\alpha}_i - e^{f_i \wdg \lambda_i} \left[e^{\mathrm{ad}(B_i)}(dF_i)\right]e^{B_i}e^{-F_i} \alpha_i  + \overset{\frown}{d\alpha}_i\\
     &= d(f_i \wdg \lambda_i) \wdg \overset{\frown}{\alpha}_i  + H_i \wdg \overset{\frown}{\alpha}_i - e^{f_i \wdg \lambda_i}
(dF_i + e_i \wdg \lambda_i) e^{B_i}e^{-F_i} \alpha_i  + \overset{\frown}{d\alpha}_i \\
     &= e_i \wdg \lambda_i \wdg \overset{\frown}{\alpha}_i - f_i \wdg \omega_i \wdg \overset{\frown}{\alpha}_i + H_i \wdg \overset{\frown}{\alpha}_i + (-)^{\overset{\frown}{\alpha}_i} e_i \wdg \pi_*\overset{\frown}{\alpha}_i - e_i \wdg \lambda_i \wdg \overset{\frown}{\alpha}_i + \overset{\frown}{d\alpha}_i \\
     &= \omega_i \wdg d\theta \wdg \overset{\frown}{\alpha}_i - \chi \wdg \overset{\frown}{\alpha}_i
     + (-)^{\overset{\frown}{\alpha}_i} e_i \wdg \pi_*\overset{\frown}{\alpha}_i + \overset{\frown}{d\alpha}_i\ .
\end{split}
\end{equation}
% \begin{equation}
%\begin{split}
%d\overset{\frown}{\alpha}_i &= -d(f_i \wdg \lambda_i) \wdg \overset{\frown}{\alpha}_i - H_i \wdg \overset{\frown}{\alpha}_i +
%e^{-f_i \wdg \lambda_i} e^{B_i} (dF_i) e^{F_i} \alpha_i  + \overset{\frown}{d\alpha}_i\\
%     &= -d(f_i \wdg \lambda_i) \wdg \overset{\frown}{\alpha}_i  - H_i \wdg \overset{\frown}{\alpha}_i + e^{-f_i \wdg \lambda_i} \left[e^{\mathrm{ad}(B_i)}(dF_i)\right]e^{B_i}e^{F_i} \alpha_i  + \overset{\frown}{d\alpha}_i\\
%     &= -d(f_i \wdg \lambda_i) \wdg \overset{\frown}{\alpha}_i  - H_i \wdg \overset{\frown}{\alpha}_i - e^{-f_i \wdg \lambda_i}
%(-dF_i - e_i \wdg \lambda_i) e^{B_i}e^{F_i} \alpha_i  + \overset{\frown}{d\alpha}_i \\
%     &= -e_i \wdg \lambda_i \wdg \overset{\frown}{\alpha}_i + f_i \wdg \omega_i \wdg \overset{\frown}{\alpha}_i - H_i \wdg \overset{\frown}{\alpha}_i + (-)^{\overset{\frown}{\alpha}_i} e_i \wdg \pi_*\overset{\frown}{\alpha}_i + e_i \wdg \lambda_i \wdg \overset{\frown}{\alpha}_i + \overset{\frown}{d\alpha}_i \\
%     &= -\omega_i \wdg d\theta \wdg \overset{\frown}{\alpha}_i + \chi \wdg \overset{\frown}{\alpha}_i
%     + (-)^{\overset{\frown}{\alpha}_i} e_i \wdg \pi_*\overset{\frown}{\alpha}_i + \overset{\frown}{d\alpha}_i\ .
% \end{split}
% \end{equation}
In both of these calculations, we have made use of Baker-Campbell-Haussdorf equations and the Clifford algebra identities discussed in section \ref{subsec:cliffalg}. Recall that these identities rely on the fact that we are restricting to forms whose Lie derivatives in the fiber directions vanish.
 
If we define
\begin{equation}
\label{twdiff3}
\mcD\beta = d\beta - (-)^\beta e \wdg \pi_*\beta - \omega \wdg d\theta \wdg \beta + \chi \wdg \beta
= d\beta - \sigma\beta + \chi\wdg\beta\ ,
\end{equation}
%\begin{equation}
%\label{twdiff3}
%\mcD\beta = d\beta - (-)^\beta e \wdg \pi_*\beta + \omega \wdg d\theta \wdg \beta - \chi \wdg \beta\ ,
%\end{equation}
we have
\begin{equation}
\mcD\overset{\frown}{\alpha}_i = \overset{\frown}{d\alpha}_i\ ,
\end{equation}
and the verification that this map induces an isomorphism follows from the invertibility of the map \eqref{hatmap3}. The expression \eqref{twdiff3} is the correct twisted cohomology for the situation where $e \wdg \omega$ only equals zero in cohomology. There is an ambiguity in the choice of $\chi$, however. It can be modified by adding any closed form. However, $\chi$ lives solely on the base, so this is just a choice of a three-form flux on the base, and choices that differ by an exact form are canonically isomorphic as in section \ref{sec:threelegs}.

Combining the results of this section, we have shown that the twisted cohomology defined by a pair $e$ and $\omega$ along with a trivialization $d\chi = e \wdg \omega$ computes the twisted cohomology $H_H^\bullet(E)$ where $E$ is the circle bundle with Euler class $e$ and $H$ is the three form given by \eqref{threeformtot}. This proves theorem 1 in the case where $d=1$.

\subsection{Higher dimensions}

To generalize this to the situation where dim $V = d > 1$, we need to ask what is the condition for a form on the base to be the push-forward of a form on the total space of a product of circle bundles. Let us choose a basis of the lattice $L$ as above and take the forms $e^a$ to define a collection of circle bundles $E_a$. Define $E$ to be the fiber product $E_1 \times_{\!\!X} \dots \times_{\!\!X} E_d$. While the choice of basis is not natural, the total space of $E$ does not depend on such a choice. This can be seen by noting that the bases of $V$ are interchanged by $\mathrm{SL}(d,\BZ)$ which also acts as the usual isomorphism of the $T^d$ fiber of $E$. We will discuss the full T-duality group later in this section.

By examining the Leray-Serre spectral sequence for the fiber bundle $E$, we see that necessary condition for a set of push-forwards $\omega_a$ to come from a three-form on the total space is that $\sum_a \omega_a \w e^a = [0]$. This is precisely the condition mentioned above. The subspace of $H^3(E)$ that has ``one leg on the fiber" is the quotient of the space of $\omega_a$ satisfying this condition by the span of collections of the form $\omega_a =C_{ab} e^b$ for any antisymmetric $C_{ab}$.

We can construct the global three form as in \eqref{threeformtot}. Let $d\chi = \sum_a \omega_a \wdg e^a$ and choose connections $\mcA^a$ on each of the $E_a$. Then, we have
\begin{equation}
\label{hdef2}
H = \sum_a \mcA^a \wdg \omega_a - \chi\ .
\end{equation}
Proceeding as above, we obtain the twisted differential operator
\begin{equation}
\label{twdiff2}
\mcD\alpha = d\alpha - \sum_a\ls (-)^\beta e^a \wdg \pi_{a\,*}\alpha + \omega_a \wdg d\theta^a \wdg \alpha\rs + \chi \wdg \alpha = d\alpha - \sigma \alpha + \chi \wdg \alpha \ .
\end{equation}
Here, $\pi_{a\,*}$ is the map from $X \times T^d$ to itself given by integrating over the $a$th $S^1$ and the pulling back to the total space, and $d\theta^a$ is the generator of the cohomology of the $a$th $S^1$. Lastly, $\sigma$ is our original Clifford algebra valued two-form which acts on forms as described in section \ref{subsec:cliffalg}. As always, we have an ismorphism between this twisted cohomology and a cohomology theory defined by forms on the $U_i$ with transitions defined analogously to the single circle case. Thus, we see that the twisted cohomology with differential given by \eqref{twdiff2} acting on $X \times T^d$ computes the twisted cohomology $H^\bullet_H(E)$ with $H$ given by \eqref{hdef2}. This proves theorem one in the case where $d$ is general, but the $\mathrm{SL}(d,\BZ)$ bundle is trivial.

\section{Combining two legs and one leg}\label{sec:alllegs}

Let us consider the case where we have both forms with two legs on the base as in section \ref{sec:twolegs}, and forms with one leg on the base as in section \ref{sec:oneleg}. Let $\sigma \in H^2(X,L)$ be a two form valued in the lattice $L \subset \cl^1$, and let $\rho \in \Omega^1(\cl^2)$ be the connection that appears in section \ref{sec:oneleg}. For simplicitly, we will assume that we can choose $\chi = 0$ as adding it to the equations is straightforward. We define the twisted differential
\begin{equation}
\mcD\alpha = d\alpha - \sigma\alpha + \rho\alpha = \mcD_\rho \alpha- \sigma\alpha\ .
\end{equation}
Then, the square of this is
\begin{equation}
\label{twoonediffsq}
\mcD^2\alpha = -\mcD_\rho(\sigma\alpha) - \sigma \mcD_\rho\alpha
\end{equation}
where $\mcD_\rho$ is the flat connection obtained from the $\Spin(d,d,\BZ)$ principal bundle as in section \ref{sec:oneleg}.
Setting this to zero, we obtain
\begin{equation}
d\sigma\alpha + \rho\sigma\alpha + \sigma d\alpha + \sigma\rho\alpha = \lp\mcD_\rho\sigma\rp\al = 0\ .
\end{equation}
To understand the final statement, recall that $\sigma$ can be thought of as a $\cl^1$ valued form and that $\cl^1$ is acted on by $\Spin(d,d)$. Thus, $\mcD_\rho$ acts on such forms as an associated connection. Note that we have used $d\sigma=0$ in the equation. This motivates us to relax our condition on $\sigma$. Choose $\sigma \in H^2_\rho(X,L)$, the cohomology of $\mcD_\rho$. Then, we have $\mcD_\rho (\sigma\alpha) = -\sigma\mcD_\rho\alpha$, and equation \eqref{twoonediffsq} is trivially satisfied. As in the beginning of the section, by making use of the inner product on $\cl^1$ multiplied by two, we obtain a map
\begin{equation}
H^2_\rho(X,L) \otimes H^2_\rho(X,L) \to H^4(X,\BZ)\ .
\end{equation}
We retain the condition that $||\sigma||^2 = [0]$, but now reinterpret it in terms of these twisted cohomology groups.

We can summarize this as follows. The most general twisted cohomology consists of the following data. We have a principal $\Spin(d,d,\BZ)$ bundle on $X$ giving rise to a flat connection $\mcD_\rho$. We have an element $\sigma \in H^2_\rho(X,L)$ in the cohomology of this flat connection which satisfies $||\sigma||^2 = [0]$. Lastly, we have an element $\gamma \in C^3(X,\BZ)$ such that $d\gamma = ||\sigma||^2$. Note that in the notation of the previous section $\chi = -\gamma$. The resulting structure depends on the addition of an exact form to $\gamma$ by a canonical isomorphism as in section \ref{sec:threelegs}, so what we have is an object in an $H^3(X,\BZ)$ torsor. As in section \ref{sec:oneleg}, we can see that these three classes and the associated conditions correspond to the non-geometric twists and Bianchi identities of \cite{Shelton:2005cf}.

\subsection{Local description}

The local description of this cohomology involves a combination of the results of the previous two sections. To begin with, we need to trivialize the form $\sigma \in H^2_\rho(X,L)$. This can be represented as a collection of forms $\sigma_i \in \Omega^2(U_i,\BZ^{2d})$ on each open set $U_i$. On the overlaps $U_{ij}$, these forms obey $\sigma_j = R_{ij}\sigma_i$ where $R_{ij}$ are the $\Spin(d,d,\BZ)$ transition functions of the bundle $E$. We can trivialize each $\sigma_i = d\tau_i$ and on the overlap obtain the ($\BZ^{2d}$-valued) one-form:
\begin{equation}
\tau_{ij} = \tau_i - R^{-1}_{ij} \tau_j\ .
\end{equation}
In addition, we have the form $d\gamma = ||\sigma||^2$. Let $\epsilon_i = \gamma - \langle\tau_i,\sigma_i\rangle$ so that $d\epsilon_i = 0$. Let $d\delta_i = \epsilon_i$. On overlaps, we make
\begin{equation}
\delta_{ij} = \delta_i - \delta_j + \langle \tau_i,\tau_{ij}\rangle\ .
\end{equation}
Notice that $d\delta_{ij}=0$. The reason for this transformation is that $\tau_i$ is akin to a connection, and the $\tau_{ij}$ is analogous to the shift one finds in transforming a connection. This is contracted with $\tau_i$ as $d\tau_i = \sigma_i$. This is closely related to the definition $B_{ij} = s^*_{ij} B_i - B_j$ of the previous section as we will see below.

We define a cohomology theory by taking forms in $\Omega^\bullet(U_i, S)$ which obey the transition function:
\begin{equation}\label{gentrans}
\alpha_j = e^{{\delta_{ij}}} R_{ij} e^{\tau_{ij}} \alpha_i
\end{equation}
where $\tau_{ij}$ acts on forms by the Clifford algebra action.

With this in hand, we form
\begin{equation}
\widehat{\alpha}_i = e^{\langle \tau_i,\tau_i\rangle}e^{ \delta_i} e^{\tau_i}  \alpha_i\ .
\end{equation}
Then,
\begin{equation}
\label{geneqver}
\begin{split}
\widehat{\alpha}_j &= e^{\langle \tau_j,\tau_j\rangle}e^{ \delta_{j}} e^{\tau_j}  \alpha_j =e^{\langle \tau_j,\tau_j\rangle} e^{ \delta_{j}}  e^{\tau_j}  e^{ \delta_{ij}} R_{ij} e^{\tau_{ij}} \alpha_i \\
	&= e^{\langle \tau_j,\tau_j\rangle}e^{\langle \tau_i, \tau_{ij}\rangle}e^{ \delta_i} R_{ij} e^{R_{ij}^{-1}\tau_j} e^{\tau_{ij}} \alpha_i = e^{\langle \tau_j,\tau_j\rangle}e^{\langle \tau_i, \tau_{ij}\rangle}e^{ \delta_i} R_{ij} e^{R_{ij}^{-1}\tau_j} e^{\tau_i - R^{-1}_{ij}\tau_j} \alpha_i \\
	&=  e^{\langle \tau_j,\tau_j\rangle}e^{\langle \tau_i, \tau_{ij}\rangle}e^{\langle \tau_{ij}, R^{-1}_{ij}\tau_{j}\rangle} e^{ \delta_i} R_{ij} e^{\tau_i}  \alpha_i = R_{ij}\widehat{\alpha}_i\ .
\end{split}
\end{equation}
Thus $\widehat{\alpha}$ is a global section of the bundle $\Omega^\bullet(X,\mcV)$. Note that we have used the invariance of the inner product under $\Spin(d,d,\BZ)$ transformations.

Taking the differential, we obtain
\begin{equation}
d\widehat{\alpha_i} =  \gamma \wdg \widehat{\alpha_i} + \sigma_i \widehat{\alpha_i} + \widehat{d\alpha_i}\ ,
\end{equation}
or
\begin{equation}
\mcD\widehat{\alpha}_i = \widehat{d\alpha_i}\ .
\end{equation}

Note that this expression is given in an open set using one of the natural trivializations. Thus, the local connection form is zero.  We could instead make a local gauge transformation, defining
\be
\widehat{\al_i}'=g_i(x)\cdot\widehat{\al_i},
\ee
in terms of which we have
\be
\label{gentwdiff}
d\widehat{\al_i}'=\lp dg_i\,g_i^{-1}\rp\w\widehat{\al_i}'+\g\w\widehat{\al_i}'+\lp g_i\s_ig_i^{-1}\rp\widehat{\al_i}'+\widehat{d\al_i}'.
\ee
%If we instead choose the connection to be of the form $dg g^{-1}$, we obtain:
%\begin{equation}
%\label{gentwdiff}
%d\widehat{\alpha_i} =  \gamma \wdg \widehat{\alpha_i} + (dg)g^{-1} \widehat{\alpha_i} + g\sigma g^{-1} \widehat{\alpha_i}+\widehat{d\alpha_i}\ .
%\end{equation}
In this case there is a local connection $-dg_i\,g_i^{-1}=g_i\,d(g_i^{-1})$, and the presence of the covariant derivative clear.

If we are in the case where the $\Spin(d,d)$ bundle, $Q$, is trivial, we can turn this into an expression in terms of untwisted forms. In particular, the triviality of the bundle implies that there exist local group valued functions $R_i : U_i \to \Spin(d,d)$ such that $R_{ij} = R^{-1}_j R_i$. Then we can form
\begin{equation}
\widehat{\alpha}_i= e^{\langle \tau_i,\tau_i\rangle}e^{ \delta_i} R_i e^{\tau_i}  \alpha_i\ ,
\end{equation}
and a slight modification of \eqref{geneqver} shows that $\widehat{\alpha}_j = \widehat{\alpha}_i$.
%\begin{equation}
%\begin{split}
%\widehat{\alpha}_j &= e^{\langle \tau_j,\tau_j\rangle}e^{ \delta_{j}}R_j e^{\tau_j}  \alpha_j =e^{\langle \tau_j,\tau_j\rangle} e^{ \delta_{j}} R_j e^{\tau_j}  e^{ \delta_{ij}} R_{ij} e^{\tau_{ij}} \alpha_i \\
%	&= e^{\langle \tau_j,\tau_j\rangle}e^{\langle \tau_i, \tau_{ij}\rangle}e^{ \delta_i} R_j R_{ij} e^{R_{ij}^{-1}\tau_j} e^{\tau_{ij}} \alpha_i = e^{\langle \tau_j,\tau_j\rangle}e^{\langle \tau_i, \tau_{ij}\rangle}e^{ \delta_i} R_i e^{R_{ij}^{-1}\tau_j} e^{\tau_i - R^{-1}_{ij}\tau_j} \alpha_i \\
%	&=  e^{\langle \tau_j,\tau_j\rangle}e^{\langle \tau_i, \tau_{ij}\rangle}e^{\langle \tau_{ij}, R^{-1}_{ij}\tau_{j}\rangle} e^{ \delta_i} R_i e^{\tau_i}  \alpha_i = \widehat{\alpha}_i\ .
%\end{split}
%\end{equation}
Thus, we have a global form on $X \times T^d$. The differential is given by the same expression as in \eqref{gentwdiff} with the $g_i$ replaced by the function $R_i$. Recall that the Lie algebra of $\Spin(d,d)$ emebedded in $\cl^2$ acts on $\cl^1$ by commutators giving rise to the fundamental representation. Thus, the adjoint action of $R_i$ on $\sigma_i$ is a rotation. This undoes the twisting of the form $\sigma_i$ by the local system $L$ and gives rise to a global form.

\subsection{T-duality}

We can now examine the action of the T-duality group $\mathrm{Spin}(d,d,\BZ)$ on this cohomology. It acts as a gauge transformation of our connection on the $\Spin(d,d)$ bundle, $Q$, taking the holonomies to their conjugates under the T-duality action. Everything is invariant under such a transformation. This can be seen explicitly as follows. If $R\in \mathrm{Spin}(d,d,\BZ)$ is the T-duality element, then we change our transition functions to $RR_{ij}R^{-1}$ which give an equivalent flat bundle. We then have an action on $\sigma$ and $\rho$ given by conjugation, \ie $\sigma \mapsto R\sigma R^{-1}$ where we regard $R$ as being an element of $\operatorname{Spin}(d,d,\Z)\subset\cl^\mathrm{ev}$. Thus, the cohomology of the T-dual situation is given by the twisted differential $\mcD^R \alpha = d\alpha + R\rho R^{-1}\alpha - R \sigma R^{-1} \alpha - \gamma \wdg \alpha$. Since the action of the T-duality group preserves the inner product $\langle \sigma,\sigma \rangle$, we do not need to modify the form $\gamma$.

As we have seen, the Clifford algebra $\cl$ acts on $S$, and the action of $\Spin(d,d,\BZ)$ is given by the exponential of elements of $\cl^2$. Furthermore, this action 
commutes with the action of the exterior differential. We can define the following map from the complex with differential $\mcD$ to that with differential $\mcD^R$ by taking $\alpha \mapsto R\alpha$. Then we have
\begin{equation}
\mcD^R R\alpha = \mcD_{\rho} R\alpha - R\sigma R^{-1} R\alpha - \gamma \wdg R\alpha = R(\mcD_\rho\alpha - \sigma \alpha - \gamma \wdg \alpha) = R \mcD\alpha\ .
\end{equation}
Furthermore, this map is invertible by applying $R^{-1}$, and the composition is precisely the identity. Thus, we see explicitly that the T-dual cohomologies are isomorphic.

\subsection{Proof of theorem 1}

It is now straightforward to prove theorem one. We will give a sketch here. Since the local system $L$ decomposes into $\widetilde{L}$ and $\widetilde{L}^*$ each associated to the $\mathrm{SL}(d,\BZ)$ bundle, $\widetilde{E}$, we can decompose $\sigma$ and $\tau$ in the basis $\eta^a$ and $\eta_a$ as in section \eqref{basisdecomp}. As discussed in the introduction, the resulting $e^a$ can be used to form a $T^d$ bundle, $T$, and the $\om_a$ a three-form, $H$, on it. The $e^a$ trivialize as $f_i^a$ and give rise to $f_{ij} = f_i - \left(\widetilde{R}^{-1}_{ij}\right)^*f_j$ where $\widetilde{R}_{ij}$ are the transition functions for $\widetilde{E}$, and we are now suppressing $L$ indices. Defining the operators $F_{ij}$ as in \eqref{firstFdef}, a global form on $T$ obeys:
\begin{equation}
\alpha_j = s^*_{ij}\alpha_i = \widetilde{R}^*_{ij} e^{-F_{ij}} \alpha_i\ ,
\end{equation}
and thus equation \eqref{buntrans} becomes
\begin{equation}
\label{geogentrans}
\alpha_j = e^{B_{ij}}s^*_{ij}\alpha_i = e^{B_{ij}}\widetilde{R}^*_{ij} e^{-F_{ij}} \alpha_i\ .
\end{equation}

We can trivialize $H_i = dB_i$ and, as before, let $B_{ij} = s^*_{ij}B_i - B_j$.  $B_i$ be written as $\delta_i + \sum_a \lambda_{i,a} \wdg d\theta^a+ \frac{1}{2} \lambda_{i,a} \wdg f^a_i$ . Since by \eqref{pullbackact}, the $\mathrm{SL}(d,\BZ)$ subgroup of $\Spin(d,d,\BZ)$ acts on forms precisely by the pullback with respect to the usual action of  $\mathrm{SL}(d,\BZ)$ on $T^d$, one can now easily verify that the transition function \eqref{gentrans} is the same as the twisted cohomology transition functions \eqref{geogentrans} (note the sign in \eqref{basisdecomp} and that $\gamma = -\chi$). We leave the case where things can be defined over the integers to the reader.

\section{Discussion}\label{sec:discuss}

The cohomology theories derived in this paper serve as a repository for the differential forms that define the total Ramond-Ramond field. However, we know that in the quantum theory the fluxes are quantized. In a geometric compactification on a geometry, $X$, without flux, the quantized fields are elements in the K-theory groups $K^0(X)$ or $K^1(X)$ for the type IIA and IIB string respectively. The Chern character gives rise to a map $\mathrm{ch} : K^\bullet(X) \to H^\bullet(X)$ where $H^\bullet(X)$ represents the sum of the even or odd cohomology groups as appropriate. This map is an isomorphism when we tensor the K-theory group by the rational numbers. In fact, the appropriate map for string theory is a combination of the A-roof genus and the Chern character.

In the presence of a three-from NSNS background $[H] \in H^3(X,\BZ)$, one can define a twisted K-theory group $K^\bullet_H(X)$ \cite{Bouwknegt:2000qt,Atiyah:2004jv} and a twisted cohomology group $H_H^\bullet(X)$ (the same as defined in section \ref{sec:threelegs}). The twisted K-theory groups serve as the repository for the quantized RR fields, and there again exists a Chern character \cite{Bouwknegt:2001vu,Atiyah:2005gu} which maps the twisted K-theory groups to the twisted cohomology groups again giving rise to an isomorphism over the rationals. In this paper, we have defined the analog of the groups $H_H^\bullet(X)$ for a choice of nongeometric flux. The goal of the sequel to this paper \cite{sequel} is to describe the analog of the group $K_H^\bullet(X)$. We provide a brief resume here.

The resemblance of the local description of these twisted theories to the transition functions of a fiber bundle is not a coincidence. It can be made precise as we will now describe. In the 1940s, Eilenberg and Steenrod presented a list of five axioms that completely characterize the cohomology $H^\bullet(X,R)$ for some ring $R$. If we drop one of these axioms (that describing the cohomology of a point), we obtain the axioms for a generalized cohomology. K-theory is an example of such a theory. The Brown representability theorem tells us that any such cohomology theory is represented by something called a spectrum. In other words, for any generalized cohomology theory $E^n(X)$, there is a sequence of spaces $E_n$ (defined up to homotopy) such that $E^n(X) \cong [X,E_n]$. Here, the notation $[X,E_n]$ is the set of homotopy classes of maps from $X$ to $E_n$. For example, for the cohomology $H^n(X,\BZ)$, the spectrum is given by the Eilenberg-MacLane spaces $K(\BZ,n)$. Similarly, for K-theory, we have that $K^0(X)$ is represented by the space of Fredholm operators, $\mcF$.

To understand what a twisting of a generalized cohomology theory is, then, notice that we can think of the space of maps between two spaces $Y\to Z$ as the space of sections of the trivial bundle $Z \times Y \to Y$. A twist of this situation is then given by a nontrivial bundle:
\begin{equation}
\begin{split}
\xymatrix{ 
Z\ar[r] & W \ar[d]^{\pi}  \\
&Y}\end{split}\ .
\end{equation}
We can then look at the homotopy classes of sections, \ie maps $s: Y \to W$ such that $\pi \circ s = \mathrm{id}$. Replacing the fiber $Z$ by a space in a spectrum defines a twisted cohomology theory.

Let us see how this works in an example \cite{Atiyah:2005gu}. Recall from section \ref{sec:threelegs} that we can describe the twisted cohomology $H^\bullet_H(X)$ by choosing an open cover $U_i$ of $X$, and defining a complex consisting of a collection of forms $\alpha_i \in \Omega^\bullet(U_i)$ that satisfy:
\begin{equation}
\label{disctrans}
\alpha_j = e^{B_{ij}} \alpha_i
\end{equation}
where $B_{ij}$ are integral closed two forms defined from $H$ as in section \ref{sec:threelegs}. On each open set, we have either even or odd degree forms, and on the transitions, we have a map which consists of wedging with a sum of forms in even degree. If we write $H\!R_\bullet$ for the spectrum corresponding to ordinary cohomology with real coefficients made 2-periodic, we have an operation $K(\BZ,2) \times H\!R_\bullet \to H\!R_\bullet$ given by the action of the exponential. From this action, we can use the transition functions \eqref{disctrans} to define a bundle
\begin{equation}
\label{discbund}
\begin{split}
\xymatrix{
H\!R_\bullet \ar[r] & W \ar[d] \\
& X}\end{split}\ .
\end{equation}
This construction can be described more conceptually as follows. $K(\BZ,2)$ is a group because we can add forms. Thus, we can form its classifying space $BK(\BZ,2) \cong K(\BZ,3)$ and the universal bundle:
\begin{equation}
\begin{split}
\xymatrix{
K(\BZ,2) \ar[r] & EK(\BZ,2) \ar[d] \\
& K(\BZ,3)}
\end{split}\ .
\end{equation}
Given the action of $K(\BZ,2)$ on $H\!R_\bullet$, we can define the associated bundle $EK(\BZ,2) \times_{K(\BZ,2)} H\!R_\bullet$. Then, given a three form represented as a map into $K(\BZ,3)$, we can pull back the bundle $EK(\BZ,2) \times_{K(\BZ,2)} H\!R_\bullet$ giving the bundle $W$ in \eqref{discbund}.

Now recall that two forms classify topological line bundles. There is an endomorphism of K-theory given by tensoring with a line bundle. This implies that there is an action $K(\BZ,2) \times \mcF \to \mcF$. The bundle $EK(\BZ,2) \times_{K(\BZ,2)} \mcF$ is then the classifying space for twisted K-theory.

Since we are dealing with torus bundles, we will define a cohomology theory by taking a space $X$ to the cohomology of the space $X \times T^d$. The spectrum for this theory is given by the $d$-fold free loop space $\mcL^dE_\bullet$. In \cite{sequel}, we will let $E_\bullet$ be the K-theory spectrum and demonstrate a homotopy action of the Clifford algebra on it. Thus, we can use the transition functions defined in this paper to define a nontrivial bundle of spectra, and the homotopy classes of sections of this bundle will define the twist of K-theory that classifies RR fluxes in backgrounds with nongeometric flux.

\section*{Acknowledgements}

We would like to thank Kevin Costello, Jacques Distler, Dan Freed, Dennis Sullivan, Brian Wecht and Timm Wrase for helpful conversations.  It is also our pleasure to thank the 5th Simons Workshop in Mathematics and Physics at Stony Brook for a stimulating environment during the completion of some of this work. A.~B. would also like to thank the Aspen Center for Physics where some of this work was completed. The research of D.~R. was supported by the National Science Foundation under Grant No. PHY-0455649. The research of A.~B. was supported by the National Science Foundation under Grant Nos. PHY-0505757 and PHY-0555575 and by Texas A\&M University. 

\bibliography{ngcoho}
\bibliographystyle{utphysnew}

\end{document}